\documentclass[onecolumn,secnumarabic,amsmath,amssymb,balancelastpage,nofootinbib]{article}

\usepackage[e]{esvect}
\usepackage{color}         
\usepackage{graphics}      
\usepackage{graphicx}      
\usepackage{epsf}          
\usepackage{bm}            

\usepackage{natbib}
\usepackage{amssymb}
\usepackage{amsmath}
\usepackage{mathrsfs}
\usepackage{framed}
\usepackage{bigints} 
\usepackage{enumitem}
\usepackage{pifont}
\usepackage[capbesideposition={left,center},facing=yes,capbesidewidth=8cm,capbesidesep=quad]{floatrow} 
\usepackage{setspace} 

\usepackage[none]{hyphenat} 

\usepackage[colorlinks=true]{hyperref}  

\definecolor{darkred}{rgb}{0.6,0,0}
\definecolor{darkgreen}{rgb}{0,0.5,0}
\definecolor{darkblue}{rgb}{0,0,0.6}
\hypersetup{ colorlinks,
linkcolor=darkblue,
filecolor=darkgreen,
urlcolor=darkgreen,
citecolor=darkred }

\begin{document}

\sloppy 

\bibliographystyle{authordate1}


\title{\vspace*{-35 pt}\Huge{Constructing and Constraining\\Wave Functions for\\Identical Quantum Particles}}
\author{Charles T. Sebens}
\date{\vspace*{-5 pt}August 30, 2016\vspace*{5 pt}\\Forthcoming in \emph{Studies in History and Philosophy of Science}\\\emph{Part B: Studies in History and Philosophy of Modern Physics}\vspace*{12 pt}}

\maketitle
\vspace*{-20 pt}
\begin{abstract}
I address the problem of explaining why wave functions for identical particles must be either symmetric or antisymmetric (the symmetry dichotomy) within two interpretations of quantum mechanics which include particles following definite trajectories in addition to, or in lieu of, the wave function: Bohmian mechanics and Newtonian quantum mechanics (a.k.a. many interacting worlds).  In both cases I argue that, if the interpretation is formulated properly, the symmetry dichotomy can be derived and need not be postulated.
\
\end{abstract}


\section{Introduction}\label{intro}

When you squeeze a marshmallow or a sponge or a down comforter it compresses easily.  Why?  Because these objects are mostly empty space.  But, when you look closely, it turns out that ordinary hard solid objects---bricks, forks, diamonds---are also mostly empty space.  In fact, they are almost entirely empty space.  All of matter is made of atoms and atoms are mostly void.  Atoms appear solid for two reasons.  First, electrons resist being squeezed toward the nucleus because it involves an increase in kinetic energy.  The lowest energy electron state balances the decrease in potential energy achieved by being near the positively charged nucleus with the increase in kinetic energy incurred by having a wave function which is tightly peaked.  The second reason for apparent solidity is that the Pauli exclusion principle forbids the electrons in more energetic states extending farther from the nucleus from being squeezed into the closer, already occupied states.

If we want to understand why objects are solid, we should seek an explanation of the Pauli exclusion principle.  The Pauli exclusion principle is a consequence of a more general principle, \emph{the spin-symmetry connection}: bosons (particles with integer spin, like photons, gluons, and mesons) have symmetric wave functions and fermions (particles with half-integer spin like protons, neutrons, electrons, and quarks) have antisymmetric wave functions.  The spin-symmetry connection is a key part of quantum mechanics and essential to many of the theory's greatest successes.  In addition to its role in the explanation of the size, structure, and solidity of atoms, the spin-symmetry connection is central to understanding more exotic phenomena such as Bose-Einstein condensates, neutron stars, lasers, and superfluid helium.  Within non-relativistic quantum mechanics, the spin-statistics connection is often taken to be an additional postulate, extremely strongly supported by empirical evidence but not derivable from more fundamental parts of the theory.

One strategy for explaining the spin-symmetry connection is to appeal to the spin-statistics theorem in relativistic quantum field theory.  The discussion here will avoid such complexities, seeking explanation within non-relativistic quantum mechanics and considering neither the connection between spin and symmetry nor the connection between symmetry and statistics.

In this article I focus on a logically weaker piece of the spin-symmetry connection, \emph{the symmetry dichotomy}: the wave function for a collection of identical particles (particles of the same type, e.g., electrons) must be either symmetric or antisymmetric.  Textbooks sometimes give a quick derivation of the symmetry dichotomy relying on the fact that the particles whose quantum states the wave function describes are identical, or at least indistinguishable (e.g. \citealp[sec. 10.3]{shankar}; \citealp[sec. 4.5]{weinberg2015}).  But, whether and in what sense there are particles distinct from the wave function is left unclear.  Here I examine the problem of explaining the symmetry dichotomy within two interpretations of quantum mechanics which clarify the connection between particles and the wave function by including particles following definite trajectories through space in addition to, or in lieu of, the wave function: (1) Bohmian mechanics and (2) a hydrodynamic interpretation that posits a multitude of quantum worlds interacting with one another, which I have called ``Newtonian quantum mechanics'' (\citealp{HDW} have called this kind of approach ``many interacting worlds'').  Versions of this second interpretation have recently been put forward by \citet{tipler,poirier2010, schiff2012,bostrom2012,bostrom2014, HDW, sebens2015}; it builds on the hydrodynamic approach to quantum mechanics (see \citealp{madelung1927,wyatt2005,holland2005}).  Bohmian mechanics and Newtonian quantum mechanics are often called ``interpretations'' of quantum mechanics, but should really be thought of as distinct physical theories which seek to explain the same body of data (those experiments whose statistics are successfully predicted by the standard methods of non-relativistic quantum mechanics).

\defcitealias{bacciagaluppi2003}{Bacciagaluppi's} 

I argue that in both interpretations the symmetry dichotomy can be derived and need not be postulated, provided the theories are formulated properly (with certain metaphysical ``links'' absent).  The two derivations are structurally similar but philosophically quite different.  I first present a derivation of the symmetry dichotomy in Bohmian mechanics which builds on \citetalias{bacciagaluppi2003} \citeyearpar{bacciagaluppi2003} illuminating analysis.  Because Bohmian mechanics without the symmetry dichotomy is a perfectly coherent theory, there is a sense in which such a derivation is not possible.  I argue that the derivation is only possible if we assume that the Bohmian guidance equation generates velocities without relying on a correspondence between particles and arguments of the wave function (the metaphysical basis for such an assumption is discussed later in section \ref{ont}), and even then the derivation is not as clean as one might like.  I then propose an explanation of the symmetry dichotomy in Newtonian quantum mechanics which parallels Bacciagaluppi's, but is simpler and stronger.  In Newtonian quantum mechanics the wave function is not part of the fundamental ontology, but simply a convenient way of describing the positions and velocities of particles inhabiting different worlds.  Because a world in Newtonian quantum mechanics is equally well represented by multiple points in configuration space, it turns out that any wave function constructed to describe these worlds will be either symmetric or antisymmetric.  I believe that the ability of Newtonian quantum mechanics to naturally explain this symmetry and others---including global phase, time reversal, and Galilean boosts \citep[sec. 6 and 12]{sebens2015}---is one of its primary virtues as an interpretation of quantum mechanics.

In section \ref{reduced} I show that the derivations from the previous sections can be recast in reduced configuration space where points related by a permutation of the locations of identical particles are identified.  This recasting of the arguments is useful in setting up the discussion of subtle metaphysical questions in the conclusion.  \citet{chen} recently proposed using the ability of various ontologies for Bohmian mechanics to explain the symmetry dichotomy to help decide between them.  In the final section I conclude that such considerations provide reason to prefer certain versions of each of these ontological accounts; for example, if one is going to take configuration space as fundamental---as in \citet{albert1996, albert2013}; \citet[sec. 4]{bostrom2014}; \citet[sec. 11, option 1]{sebens2015}---it should be the reduced and not the full configuration space.

I am concerned primarily with the question of why identical particles behave differently from non-identical ones, not the question of why we don't observe them behaving in a more exotic way: as either paraparticles or anyons.  Paraparticles are automatically forbidden by the wave function formalism used here and will only be mentioned briefly.  Anyons are also forbidden by the formalism, though I will explain in section \ref{reduced} why even if we alter the formalism to allow them (by permitting multi-valued wave functions) they cannot exist in three or more spatial dimensions (drawing insight from \citealp{leinaas1977} but presenting a distinct argument).

\section{Explaining the Symmetry Dichotomy: Bohmian Mechanics}\label{explainingBM}

Let's first consider the problem of finding an explanation for the symmetry dichotomy in the context of Bohmian mechanics.  For an introduction to the theory, see \citet{berndl1995, sep-qm-bohm}.  According to Bohmian mechanics, the wave function $\Psi$ for multiple particles (ignoring spin) obeys the Schr\"{o}dinger equation
\begin{equation}
i\hbar\frac{\partial}{\partial t}\Psi(\vec{x}_1,\vec{x}_2,...,t)
=\left(\sum_k{\frac{-\hbar^2}{2 m_k}\nabla_k^2}+V(\vec{x}_1,\vec{x}_2,...,t)\right)\Psi(\vec{x}_1,\vec{x}_2,...,t)\ ,
\label{schrod}
\end{equation}
where $\Psi$ is a complex-valued function of particle configuration $(\vec{x}_1,\vec{x}_2,...)$ and time $t$, $m_k$ is the mass of particle $k$, $\nabla_k^2$ is the Laplacian with respect to the $k$-th  (vector) argument of the function it acts on, and $V$ is the classical potential energy of particle configuration $(\vec{x}_1,\vec{x}_2,...)$ at $t$. In addition to the wave function, there are particles which at all times have well-defined locations.  There is a second dynamical law which is introduced to govern the motions of the particles, the guidance equation.  If we write the wave function in polar form as $\Psi(\vec{x}_1,\vec{x}_2,...,t)=R(\vec{x}_1,\vec{x}_2,...,t)e^{i\theta(\vec{x}_1,\vec{x}_2,...,t)}$---where $R$ assigns a positive real number to each possible configuration and $\theta$ assigns an angle (phase)---the velocity of particle $k$ at $t$ is given by
\begin{equation}
\vec{v}_k(t)=\frac{\hbar}{m_k}\vec{\nabla}_k\theta(\vec{x}_1,\vec{x}_2,...,t)\ .
\label{guideq2}
\end{equation}

Consider a collection of particles which have the same intrinsic properties: mass, charge, spin, etc.  That is, consider a collection of electrons, a collection of positrons, a collection of photons, or some other collection of particles of the same type.  I will call these identical particles, but all I mean by that is that they are particles of the same type.  Since the particles follow distinct paths through space, they cannot be truly identical to one another and will be distinguishable (at least from a God's eye point of view).  Associated with these $N$ particles is a wave function which assigns a complex number to each possible arrangement of the particles, $\Psi(\vec{x}_1,\vec{x}_2, \vec{x}_3,...,\vec{x}_N,t)$, where $\vec{x}_1$ gives a location of one of the particles, $\vec{x}_2$ of another, etc.  The wave function assigns a complex number to every point in $3N$ dimensional configuration space, where a point in configuration space is picked out by giving the location of each of the $N$ particles in 3 dimensional space: $(\vec{x}_1,\vec{x}_2, \vec{x}_3,...,\vec{x}_N)$.  Two wave functions are considered equivalent if and only if they differ by at most a global phase.  If the wave function always assigns the same complex number to any pair of points in configuration space related by a swap of two particle positions (a permutation), it is \emph{symmetric}:
\begin{equation}
\forall i,j\ \Psi(\vec{x}_1,...,\vec{x}_i,...,\vec{x}_j,...,\vec{x}_N,t)=\Psi(\vec{x}_1,...,\vec{x}_j,...,\vec{x}_i,...,\vec{x}_N,t)\ .
\label{symcon}
\end{equation}
Alternatively, the wave function may be such that the values at two such points differ by a phase shift of $\pi$, in which case the wave function is \emph{antisymmetric}:
\begin{equation}
\forall i,j\ \Psi(\vec{x}_1,...,\vec{x}_i,...,\vec{x}_j,...,\vec{x}_N,t)=-\Psi(\vec{x}_1,...,\vec{x}_j,...,\vec{x}_i,...,\vec{x}_N,t)\ .
\label{asymcon}
\end{equation}
General complex-valued functions of $N$ vector arguments will be neither symmetric nor antisymmetric, they will be \emph{asymmetric}.  The symmetry dichotomy is the claim that asymmetric states are forbidden.  If we assume that $V$ is invariant under permutation of the particle locations it takes as arguments, then the Hamiltonian in \eqref{schrod} acts symmetrically\footnote{To say the Hamiltonian ``acts symmetrically'' is just to say that it commutes with any permutation operator (these operators are discussed at the end of the section).} on the wave function.  A symmetric(antisymmetric) wave function will remain symmetric(antisymmetric) over time.

Before examining why the symmetry dichotomy holds in Bohmian mechanics, let's consider a brief parable.  In classical mechanics the net force $\vec{F}_k$ on a body $k$ with inertial mass $m_{I_k}$ results in an acceleration of $\vec{F}_k/m_{I_k}$.  This is expressed by Newton's second law,
\begin{equation}
\vec{F}_k=m_{I_k}\vec{a}
\label{secondlaw}
\end{equation}
If the only relevant force is gravity, the net force can be found by summing the gravitational attraction from all other bodies (indexed by $i$),
\begin{equation}
\vec{F}_k=\sum_i{G\frac{m_{G_k}m_{G_i}}{|\vec{r}_{ki}|^2}\hat{r}_{ki}}
\label{gravity}
\end{equation}
Here $G$ is the gravitational constant, $\vec{r}_{ki}=\vec{x}_i-\vec{x}_k$, $\hat{r}_{ki}=\vec{r}_{ki}/|\vec{r}_{ki}|$, and $m_{G_i}$ is the gravitational mass of body $i$.  As formulated here, these laws allow for each body to have a distinct gravitational and inertial mass.  It is possible that for some bodies, $m_{G_k}\neq m_{I_k}$.  As it turns out, no body exercises that freedom.  The gravitational mass is, as far as we can tell (and we are looking\footnote{See the E\"{o}tv\"{o}s experiment and recent variations on it.}), always exactly equal to the inertial mass.  We can remove this unnecessary freedom in specifying the state of a system by imposing the requirement that $m_{G_k}$ and $m_{I_k}$ always be equal as an additional postulate.  Alternatively, we can reformulate the theory so that the equality of the masses falls out automatically---as is trivially accomplished by replacing the two kinds of mass in \eqref{secondlaw} and \eqref{gravity} with a single mass $m_k$ for each body $k$.

Similarly, one can coherently formulate Bohmian mechanics without the symmetry dichotomy (below), allowing more freedom in the specification of a quantum state than nature realizes.  There is thus a sense in which proving or deriving the dichotomy is impossible.  Any supposed derivation must include some form of magic, be it overt or covert.  We can, of course, forego hopes of derivation and instead impose the symmetry dichotomy as an additional law or postulate analogous to adding the requirement that gravitational and inertial mass never differ (e.g., \citealp[ch. 14]{messiah1966}; \citealp[sec. 17.3]{ballentine}; \citealp[sec. 13.3]{gasiorowicz}; \citealp[sec. 5.1.1]{griffiths}; \citealp[sec. 7.2]{sakurai2011}).  Even though a certain kind of proof is impossible, an explanation of the dichotomy can be given if the theory is properly formulated.  In the version of Bohmian mechanics I will call ``unlinked Bohmian mechanics'' the dichotomy is automatically satisfied (just as the equality of gravitational and inertial mass followed automatically from the one-kind-of-mass formulation of classical mechanics).

Suppose the symmetry dichotomy is not imposed and identical particles are permitted to have wave functions which are neither symmetric nor antisymmetric, just like non-identical particles.  In such a theory, the value of the wave function for a collection of identical particles at $(\vec{x}_1,\vec{x}_2, \vec{x}_3,...)$ will generally differ from the value at $(\vec{x}_2,\vec{x}_1, \vec{x}_3,...)$, where the first two positions have been permuted, by more than just a phase.  As a consequence, the probability assigned to these two arrangements of particles---to these two points in configuration space---will in general be different.  As these two arrangements of particles will look the same, one calculates the probability of seeing that kind of arrangement by summing over the different permutations (this is unproblematic and what is done already if the wave function on configuration space is normalized in a standard way,\footnote{That is, if it is normalized such that integrating $|\Psi|^2$ over all of configuration space gives one.} \citealp[pg. 584]{messiah1966}).  Particle velocities can be calculated perfectly well using the guidance equation  \eqref{guideq2}.  Thus, the symmetry dichotomy is not a consequence of the fact that the particles are identical.  The theory just described would do well at predicting some possible observations, but not what we've seen in our laboratories.  It would be mysterious why identical particles are always in symmetric or antisymmetric states and thus mysterious why electrons in atoms always obey the Pauli exclusion principle, why neutron stars don't implode, why superfluid helium has zero viscosity, etc. (just like classical mechanics with unconstrained gravitational and inertial masses would leave it mysterious why the gravitational and inertial masses always happen to be the same).

There is an alternative way of formulating Bohmian mechanics for identical particles from which the symmetry dichotomy can be derived.  This reformulation requires modifying the connection between particles and wave as encoded in the guidance equation.  Suppose that there is no correspondence between the particles and the dimensions of the configuration space on which the wave function is defined.  The guidance equation \eqref{guideq2} must be revised because it relied on this correspondence---it relates the velocity of $k$-th particle to the $k$-th gradient of the phase.  The guidance equation must somehow assign a velocity to each of the $N$ particles given just the wave function and the fact that there is a particle at $\vec{x}_1$, another of the same kind at $\vec{x}_2$, another at $\vec{x}_3$, etc.  One way to do this is to say that the velocity must be independent of the way the particles are mapped to coordinates in configuration space, satisfying \eqref{guideq2} for any of the $N!$ mappings.  The velocity of the particle at $\vec{x}_1$ would then be calculable in many ways,
\begin{align}
\vec{v}_{\vec{x}_1}(t)&=\frac{\hbar}{m}\vec{\nabla}_1\theta(\vec{x}_1,\vec{x}_2,\vec{x}_3,...,t)
\nonumber
\\
&=\frac{\hbar}{m}\vec{\nabla}_2\theta(\vec{x}_2,\vec{x}_1,\vec{x}_3,...,t)
\nonumber
\\
&=\frac{\hbar}{m}\vec{\nabla}_3\theta(\vec{x}_2,\vec{x}_3,\vec{x}_1,...,t)
\nonumber
\\
&\quad\quad\quad\vdots
\ \ \ \quad\quad\quad\quad\quad\quad\quad\ .
\label{guideqred}
\end{align}
In the second line, $\vec{\nabla}_2\theta(\vec{x}_2,\vec{x}_1,\vec{x}_3,...,t)$ is the derivative with respect to its second argument, $\vec{x}_1$.  To be a little more careful, we can write this as $\vec{\nabla}_{\vec{z}}\:\theta(\vec{x}_2,\vec{z},\vec{x}_3,...,t)\big|_{\vec{z}=\vec{x}_1}$, where ``$\big|_{\vec{z}=\vec{x}_1}$'' indicates that the gradient is evaluated at $\vec{x}_1$.  It will be useful to introduce an alternative notation ``$\vec{\nabla}_{\vec{x}_1}$'' which indicates the gradient with respect to whichever argument $\vec{x}_1$ sits in, evaluated at $\vec{x}_1$.  The same gradient can then be written as $\vec{\nabla}_{\vec{x}_1}\theta(\vec{x}_2,\vec{x}_1,\vec{x}_3,...,t)$.

For \eqref{guideqred} to yield a well-defined velocity---and to avoid inconsistency---the different lines of the above equation must agree.  That is, in any physical history allowed by the laws these lines must always agree.  It is this requirement that will generate the symmetry dichotomy.\footnote{Instead of taking the guidance equation \eqref{guideq2} as the dynamical law governing the motion of particles in Bohmian mechanics, one might prefer to take the second order equation giving each particle's acceleration as the dynamical law and treat \eqref{guideq2} as a constraint on initial conditions which, because of the nature of the dynamics, will then hold at all times (as in \citealp[sec. 4]{bohm1952}).  On this understanding of the theory, instead of modifying the guidance equation as in \eqref{guideqred}, it would be natural to modify the second order equation so that the acceleration of the particle does not rely on a particular connection between particles and arguments of the wave function.  I can then see two possible strategies for deriving the symmetry dichotomy.  First, one could also modify the constraint on initial velocities as in \eqref{guideqred}.  With both modifications in place, \eqref{guideqred} will hold at all times and the derivation can proceed as described in this section.  Second, one could argue from the modified second order equation of motion alone that the magnitude $R$ of the wave function must be symmetric (see footnote \ref{troubletroubletrouble} and \citealp{bacciagaluppi2003}) and then argue from this that the wave function must be either symmetric or antisymmetric (as outlined in footnote \ref{probstrategy}).}  Let's call this modified version of Bohmian mechanics where the connection between particle and wave is weakened \emph{unlinked Bohmian mechanics} and the original version \emph{linked Bohmian mechanics}.  Exactly what this ``link'' might be (metaphysically) and how this distinction relates to debates about the ontology of Bohmian mechanics will be discussed in section \ref{ont}.

The explanation of the symmetry dichotomy in unlinked Bohmian mechanics will proceed in three steps, using unlinked Bohmian mechanics to provide a foundation for the derivation in \citet{bacciagaluppi2003} (and noting deficiencies in the derivation, so understood).  First step: showing that the gradients of the phase $\theta$ are symmetric under permutation.  Second step: deriving from this that the phase differs by an integer multiple of $\pi$ at points of configuration space related by a permutation of particle locations.  Third step: arguing that the magnitude of the wave function is also symmetric under permutation.

\textbf{Step 1: Symmetry of $\bm{\vec{\nabla}\theta}$}\ \ \  Enforcing consistency between the first two lines of \eqref{guideqred} yields the requirement that
\begin{equation}
\vec{\nabla}_1\theta(\vec{x}_1,\vec{x}_2,\vec{x}_3,...,t)=\vec{\nabla}_2\theta(\vec{x}_2,\vec{x}_1,\vec{x}_3,...,t)\ .
\label{agreement}
\end{equation}
In the new notation introduced above, we can rewrite this as,
\begin{equation}
\vec{\nabla}_{\vec{x}_1}\theta(\vec{x}_1,\vec{x}_2,\vec{x}_3,...,t)=\vec{\nabla}_{\vec{x}_1}\theta(\vec{x}_2,\vec{x}_1,\vec{x}_3,...,t)\ .
\label{agreement2}
\end{equation}
\citet{bacciagaluppi2003} motivates this key constraint as follows: ``the velocity ... of particle 1 in a given configuration is equal to that of particle 2 in the configuration with the particles exchanged. This is the natural requirement of \emph{indistinguishability} at the level of particle trajectories.''  I take unlinked Bohmian mechanics to provide an explanation of why we might expect this critical ``requirement of indistinguishability'' to hold.  The intelligibility of linked Bohmian mechanics shows that this requirement is not a trivial feature of the Bohmian mechanics of identical particles.

As each particle must have a unique velocity, we can generalize \eqref{agreement2} to
\begin{equation}
\forall i,\ \ \vec{\nabla}_{\vec{x}_i}\theta(\vec{x}_1,\vec{x}_2,\vec{x}_3,...,t)=\vec{\nabla}_{\vec{x}_i}\theta(\vec{x}_2,\vec{x}_1,\vec{x}_3,...,t)\ .
\label{foralliagreement}
\end{equation}

\textbf{Step 2: $\bm{\alpha=n\pi}$}\ \ \   For this step it is necessary to assume that \eqref{foralliagreement} does not just hold for the actual arrangement of particles, but for other possible arrangements as well.  However, it is not obvious why the condition should be extended.  According to Bohmian mechanics, there really are particles with particular locations and velocities.  If those velocities are to obey the guidance equation for unlinked Bohmian mechanics, \eqref{foralliagreement} must hold for the actual particle locations at all times.  But, why must it hold for other possible locations?  We cannot allow \emph{combinations} of wave function and particle configuration for which the guidance equation does not generate well-defined particle velocities, but why must we reject combinations where the velocity \emph{would be} undefined \emph{were the} particles in different locations?  That is, why must we reject as physically impossible (as forbidden by the theory) the apparently possible physical states (of particles and wave) where \eqref{foralliagreement} holds across the $N!$ points in configuration space that accurately represent the true particle positions but does not hold everywhere throughout the vast reaches of configuration space?

Ultimately, I believe the danger from violations of \eqref{foralliagreement} is indirect.  Typically, wave functions that fail to satisfy \eqref{foralliagreement} somewhere in configuration space will quickly fail to satisfy it almost everywhere, including whichever points give the actual particle locations.  Wave functions which violate \eqref{foralliagreement} at points in configuration space that don't represent the actual configuration are thus ruled out not because they are in violation of the laws, but because they \emph{will} lead to violation of the laws in the future.  This is legitimate reasoning as the laws must be obeyed everywhere and always.  So, these wave functions really are forbidden.  However, the way in which these wave functions are forbidden is somewhat odd.  In ordinary physical theories states which are allowed by the laws do not evolve into ones which are in violation of them.  I take this oddity to be a disadvantage of the explanation in unlinked Bohmian mechanics as compared to explanation offered by Newtonian quantum mechanics (to be discussed in the following section): the Bohmian explanation is temporally non-local.  Also, note that the above reasoning may not rule out all \eqref{foralliagreement} violating wave functions: there may be some mischievous wave functions which manage to always satisfy \eqref{foralliagreement} at the actual configuration while violating it elsewhere in configuration space.\footnote{It is not clear how much of a problem it would be if the Bohmian derivation permitted a few rare asymmetric wave functions.  My feeling is that it would be empirically fine (you wouldn't expect to see these weird wave functions), but unsettling from a foundational perspective.}  Despite these concerns, let us proceed under the assumption that \eqref{foralliagreement} is satisfied for merely possible particle configurations as well as the actual one.

Still, we cannot assume that \eqref{foralliagreement} holds absolutely everywhere in configuration space since---even in linked Bohmian mechanics---the velocities particles would have if they were in a particular configuration are not defined everywhere in configuration space by the guidance equation.  In particular, they are not defined where the wave function is zero (as the phase of the wave function is undefined).  If $\theta$ and its gradients were defined everywhere, we could straightforwardly conclude that $\theta(\vec{x}_1,\vec{x}_2,\vec{x}_3,...,t)=\theta(\vec{x}_2,\vec{x}_1,\vec{x}_3,...,t)$ from the fact that the gradients agree everywhere \eqref{foralliagreement} and that the equality must hold when $\vec{x}_1=\vec{x}_2$.  If the region on which $R \neq 0$ is connected, it follows from the fact that the gradients with respect to every position vector must agree that the phase can only differ by at most a global additive constant between locations in configuration space that differ by a swap of the first and second argument,
\begin{equation}
\theta(\vec{x}_1,\vec{x}_2,\vec{x}_3,...,t)=\theta(\vec{x}_2,\vec{x}_1,\vec{x}_3,...,t)-\alpha\ .
\label{alphaphase2}
\end{equation}
Since \eqref{alphaphase2} holds wherever $\theta$ is defined, the value of $\theta$ at a point where the locations $\vec{x}_1$ and $\vec{x}_2$ are permuted is always $\alpha$ more than its value at the original point.  As two permutations return you to where you started,
\begin{equation}
\theta(\vec{x}_1,\vec{x}_2,\vec{x}_3,...,t)=\theta(\vec{x}_1,\vec{x}_2,\vec{x}_3,...,t)-2\alpha\ ,
\label{alphaphase17}
\end{equation}
$\alpha$ must be either $0$ or $\pi$.\footnote{Here is another way to see that alpha in \eqref{alphaphase2} is an integer multiple of $\pi$ which more closely parallels the explanation in section \ref{reduced}.  Consider integrating the gradients of $\theta$ along a path $C$ in configuration space from the point $(\vec{x}_1,\vec{x}_2,\vec{x}_3,...)$ to the point $(\vec{x}_2,\vec{x}_1,\vec{x}_3,...)$ where $\vec{x}_1$ and $\vec{x}_2$ have been permuted and call the result $\alpha$,
\begin{equation}
\int_C{\left\{\sum_i{\left[\vec{\nabla}_{i}\theta \cdot d\vec{\ell}_i\right]}\right\}}=\alpha\ .
\end{equation}
This might be, for example, the path of the solid arrow connecting the cross and the dot in figure \ref{2P2D}.  Next consider a return path $C'$ which proceeds not back along $C$, but instead along the path generated by permuting each point in $C$ (as in the dotted path in figure \ref{2P2D}).  Integrating the gradients of theta along $C'$ must give the same contribution $\alpha$ since each $\vec{\nabla}_{i}\theta$ is symmetric under the permutation of $\vec{x}_1$ and $\vec{x}_2$.  But, the path formed by fusing $C$ and $C'$ is a closed loop which returns us to our starting point and integrating along such a path must give an integer multiple of $2\pi$ (as $\theta$ is single-valued).  Thus, $2\alpha=2\pi n$.\label{earliermess}}

One can easily construct troublesome states where the $R \neq 0$ region is not connected.  For example, consider any real-valued antisymmetric wave function, which must be positive some places and negative others---to be antisymmetric---and yet its phase cannot move continuously from $0$ to $\pi$ as it never takes intermediate values.  Since the permutation symmetry of the Hamiltonian entails that a symmetric(antisymmetric) wave function will always be symmetric(antisymmetric), it would suffice to find a time when the $R\neq 0$ region is connected and demonstrate that at that time the wave function must be symmetric(antisymmetric).  Bacciagaluppi's proof thus requires\footnote{\citet[\textsection 3]{bacciagaluppi2003} acknowledges this requirement.  Bacciagaluppi argues that we can exclude problematic wave functions by varying the potential $V$ in \eqref{schrod} and requiring that the velocities would be well-defined even if the potential were different.  He uses this trick again in step 3.} that there is a time when the region of configuration space in which the wave function is non-zero is connected.\footnote{Actually, a weaker condition may be sufficient.  It follows from \eqref{schrod} that
\begin{equation}
\frac{\partial \theta}{\partial t}=\sum_i{\left\{\frac{\hbar}{2 m}\frac{\nabla_i^2 R}{R}-\frac{m}{2 \hbar}|\vv{v}_i|^2\right\}}-\frac{V}{\hbar}\ ;
\label{thetaT}
\end{equation}
see \citet[eq. 17]{bohm1952}; \citet[eq. 7]{bacciagaluppi2003}; \citet[eq. 14]{sebens2015}.  Thus the time derivative of theta is fixed by the particle velocities and $R$.  If we require that the magnitude of $\Psi$ agree at two permuted points \eqref{realparts}, then $\frac{\partial\theta}{\partial t}$ must be the same at $(\vec{x}_1,\vec{x}_2,\vec{x}_3,...,t)$ and $(\vec{x}_2,\vec{x}_1,\vec{x}_3,...,t)$.  Since the derivative of $\theta(\vec{x}_1,\vec{x}_2,\vec{x}_3,...,t)$ and $\theta(\vec{x}_2,\vec{x}_1,\vec{x}_3,...,t)$ with respect to every argument (including time) must agree, all that is required is that there is some period of time over which the region of configuration-space-time for which $R\neq 0$ is connected.  Analogous reasoning applies to the case of Newtonian quantum mechanics.  I am uncertain whether realistic cases satisfy either the stronger requirement in the main text or the weaker requirement discussed here, though unrealistic cases with everlasting infinite potentials will clearly cause trouble for both.\label{potato}}  Let us assume that there is.

\textbf{Step 3: Symmetry of $\bm{R}$}\ \ \   There are two routes to showing that $R$ is symmetric.  The first route appeals to probabilities.  If we assume that the probability, $R^2$, for the particles being in the configuration $(\vec{x}_1,\vec{x}_2,\vec{x}_3,...)$ must be the same as that for $(\vec{x}_2,\vec{x}_1,\vec{x}_3,...)$, then it follows immediately that
\begin{equation}
R(\vec{x}_1,\vec{x}_2,\vec{x}_3,...,t)=R(\vec{x}_2,\vec{x}_1,\vec{x}_3,...,t)\ .
\label{realparts}
\end{equation}
The idea behind this requirement is that probabilities are being assigned to the very same arrangement of particles and thus cannot differ.  What exactly would we take to be more probable than what if $R(\vec{x}_1,\vec{x}_2,\vec{x}_3,...,t)$ was greater than $R(\vec{x}_2,\vec{x}_1,\vec{x}_3,...,t)$?  However, such concerns are far from decisive (though textbooks sometimes appear to find them so, e.g., \citealp[sec. 9.4]{ohanian}).  As was discussed when introducing linked Bohmian mechanics, we can simply say that the probability of finding the particles in this arrangement is the sum of $R(\vec{x}_1,\vec{x}_2,\vec{x}_3,...,t)^2$ and $R(\vec{x}_2,\vec{x}_1,\vec{x}_3,...,t)^2$ and $R^2$ for all other points in configuration space reached by permuting the particle positions.  The terms in the sum need not themselves correspond to probabilities of anything in particular and need not be equal.

The second route to \eqref{realparts} appeals again to the fact that the velocities must be well-defined.  (This is the route taken by \citealp{bacciagaluppi2003}.)  If the magnitude of the wave function is not symmetric, this will tend to destroy the symmetry of the gradients of $\theta$ because the time evolution of the wave function depends on both its magnitude and phase.  By examining how the phase would change over time in accordance with the Schr\"{o}dinger equation \eqref{schrod}, \citet[sec. 3]{bacciagaluppi2003} gives a complex proof (which I will not reproduce here) that $R$ must be symmetric if $\eqref{foralliagreement}$ is to hold at all times no matter the potential $V$.\footnote{One can see that an asymmetric $R$ would be problematic by looking at \eqref{thetaT} in footnote \ref{potato}.  If the potential $V$ and the velocities $\vv{v}_i$ are symmetric, then any asymmetry in $R$ which leads to an asymmetry in $\sum_i{\left\{\frac{\nabla_i^2 R}{R}\right\}}$ will cause the phase to become asymmetric.\label{troubletroubletrouble}}  Though the details of the argument are complicated, the result should seem plausible.  Just as allowing the gradients of the phase at permuted points that don't represent the actual configuration to disagree will lead to trouble for the actual velocities being well-defined in unlinked Bohmian mechanics, so will allowing $R$ to not be symmetric.\footnote{If (unlike me) you think that considerations of probability necessitate that $R$ must be symmetric at all times in unlinked Bohmian mechanics, you could alternatively take this as the starting point for the derivation of the symmetry dichotomy (as opposed to the modified guidance equation).  The intermingling of phase and magnitude just mentioned could be used to argue that the gradients of $\theta$ must be symmetric, as derived in step 1.  Asymmetric gradients of $\theta$ would be forbidden because they would generally lead to an asymmetric $R$ as $R$ evolves via
\begin{equation}
\frac{\partial R}{\partial t}=\frac{-1}{2R}\sum_i{\frac{\hbar}{m_i} \vec{\nabla}_i \cdot \Big(R^2\: \vec{\nabla}_i\theta\Big)}\ ,
\end{equation}
which is derivable from the Schr\"{o}dinger equation like \eqref{thetaT}.  Step 2 could then proceed as before.\label{probstrategy}}   This is a second point where the explanation offered in unlinked Bohmian mechanics is temporally non-local: wave functions violating \eqref{realparts} may not be breaking any laws at that moment, but such momentary indiscretion will typically lead to future unlawful behavior.  In the next section we will see that Newtonian quantum mechanics does not need to appeal to merely possible configurations or potentials and can provide a cleaner, quicker, and temporally local derivation of \eqref{realparts}.

Since $R$ must be symmetric under the permutation of the first two arguments (step 3) and $\theta$ must shift by a multiple of $\pi$ under permutation (step 2), it follows that the wave function must be either symmetric or antisymmetric under the permutation of the first two arguments.  As the decision to permute the first two arguments was arbitrary, the wave function must be symmetric or antisymmetric under any permutation of two particle positions.

As of now it appears to be an open possibility that the wave function might be symmetric with respect to some permutations and antisymmetric with respect to others.  In fact, such a mix of symmetries is not possible (\citealp[pg. 396, fn. 2]{blokhintsevQM}; \citealp[sec. 11.4]{BvF}; \citealp[sec. 4]{bacciagaluppi2003}).

Define the permutation operator $\widehat{P}_{ij}$ as swapping the $i$-th and $j$-th vector arguments of the function it acts on.  Consider two arbitrary permutations $\widehat{P}_{ij}$ and $\widehat{P}_{mn}$ such that
\begin{align}
\widehat{P}_{ij}\Psi(\vec{x}_1,...,\vec{x}_i,...,\vec{x}_j,...,t)&=\Psi(\vec{x}_1,...,\vec{x}_j,...,\vec{x}_i,...,t)
\nonumber
\\
&=e^{i \alpha_{ij}}\Psi(\vec{x}_1,...,\vec{x}_i,...,\vec{x}_j,...,t)\ .
\label{swapping1}
\end{align}
and similarly for $\widehat{P}_{mn}$, where each alpha is an integer multiple of $\pi$; that is, where $\Psi$ is either symmetric or antisymmetric under each permutation.  The swapping of $\vec{x}_i$ and $\vec{x}_j$ can be enacted simply by the operator $\widehat{P}_{ij}$ or circuitously via $\widehat{P}_{mj}\widehat{P}_{ni}\widehat{P}_{mn}\widehat{P}_{ni}\widehat{P}_{mj}$.  Acting on the wave function with this second string of operators yields
\begin{align}
&\widehat{P}_{mj}\widehat{P}_{ni}\widehat{P}_{mn}\widehat{P}_{ni}\widehat{P}_{mj}\Psi(\vec{x}_1,...,\vec{x}_i,...,\vec{x}_j,...,t)
\nonumber
\\
&\quad\quad\quad\quad\quad=e^{2 i \alpha_{mj}}e^{2 i \alpha_{ni}}e^{i \alpha_{mn}}\Psi(\vec{x}_1,...,\vec{x}_i,...,\vec{x}_j,...,t)
\nonumber
\\
&\quad\quad\quad\quad\quad= e^{i \alpha_{mn}}\Psi(\vec{x}_1,...,\vec{x}_i,...,\vec{x}_j,...,t)
\ .
\label{swapping2}
\end{align}
Since the result of the operation in \eqref{swapping1} and \eqref{swapping2} must be the same, $e^{i \alpha_{ij}}=e^{i \alpha_{mn}}$ and thus either both permutations flip the sign of the wave function or neither do.  Since this holds for every pair of permutations, the wave function must be either symmetric for all permutations or antisymmetric for all permutations.

At this point it is worth noting that the above derivation of the symmetry dichotomy in Bohmian mechanics does not explain why identical particles do not behave like anyons or paraparticles---it utilizes a framework in which such behavior is impossible.  We could have alternatively started with a wider space of possible quantum states.  For example, one might permit the wave function to be multi-valued.  This would allow for wave functions which describe anyons,  satisfying \eqref{alphaphase2} and \eqref{realparts} but neither \eqref{symcon} nor \eqref{asymcon} as $\alpha$ is not an integer multiple of $\pi$ (\citealp[\textsection 3.2]{leinaas1977}; \citealp{wilczek1982}; \citealp[ch. 3]{morandi1992}).  As none of the fundamental particles are anyons, one should seek to remove this possibility.  Fortunately, by using single-valued wave functions we have already forbidden such states.  (An arguably deeper explanation as to why they're forbidden, appealing to the single-valued nature of the gradient of the phase as opposed to the wave function itself, is given in section \ref{reduced}.) Similarly, the formalism used here already rules out the possibility of paraparticles \citep[\textsection 2]{baker2014}.  Here quantum states are taken to be given by collections of complex-valued functions which differ by at most a global phase factor whereas paraparticle states would be represented by larger collections of functions.  Though it may be possible to construct paraparticle theories which are consistent with the data, one need not allow them and forbidding them leaves us with fewer possible states.  Just as Newtonian gravity with one kind of mass automatically rules out disagreement between gravitational and inertial mass, the wave function formalism used here conveniently rules out anyons and paraparticles.

\section{Explaining the Symmetry Dichotomy: Newtonian Quantum Mechanics}\label{explainingNM}

Let us now consider the problem of explaining the symmetry dichotomy from the perspective of Newtonian quantum mechanics, a close relative of Bohmian mechanics.  According to Newtonian quantum mechanics, at the fundamental level there are just particles with definite positions and velocities---no wave function.  These particles are grouped into worlds (all of which are taken to have the same number of each type of particle\footnote{Though this would presumably not hold in a relativistic version of the theory where particles can be created and destroyed (unless the legendary Dirac sea is real).}).  Each world can be represented by a point moving through configuration space.  The flow of this large collection of worlds in configuration space is described (at a coarse-grained level\footnote{As stated here, the theory only gives the dynamics of worlds at a coarse-grained level where they are described by a density and velocity field which incompletely specify their precise states (just like one might describe a fluid in terms of a density and velocity field).  The density $\rho(\vec{x}_1,\vec{x}_2,...,t)$ gives the average number of worlds per volume of configuration space around the point $(\vec{x}_1,\vec{x}_2,...)$ at $t$ and $\vec{v}_i(\vec{x}_1,\vec{x}_2,...,t)$ gives the average velocity of the particle whose location is given by the $i$-th argument around that point in configuration space (see \citealp[sec. 5]{sebens2015}; this is similar to the way fluids are coarse-grained in, e.g., \citealp[sec. 2.2]{chapman1970}).  There is assumed to be some ``microdynamical,'' fine-grained story about how the individual worlds are moving and interacting which justifies this coarse-grained description (see \citealp{HDW} for progress on this front).}) by a density of worlds $\rho(\vec{x}_1,\vec{x}_2,...,t)$ (normalized so that integrating it over all of configuration space gives $1$) and a velocity field $\vec{v}_i(\vec{x}_1,\vec{x}_2,...,t)$ for each particle $i$.  As worlds are neither created nor destroyed, the flow obeys a continuity equation
\begin{equation}
\frac{\partial\rho(\vec{x}_1,\vec{x}_2,...,t)}{\partial t}=-\sum_i{\vec{\nabla}_i \cdot \Big(\rho(\vec{x}_1,\vec{x}_2,...,t)\vec{v}_i(\vec{x}_1,\vec{x}_2,...,t)\Big)}\ .
\label{hamster}
\end{equation}
The dynamical evolution of the velocity fields is given by a Newtonian force law
\begin{equation}
m_k \vec{a}_k(\vec{x}_1,\vec{x}_2,...,t)=-\vec{\nabla}_k\left[\sum_i \frac{-\hbar^2}{2 m_i}\left(\frac{\nabla_i^2 \sqrt{\rho(\vec{x}_1,\vec{x}_2,...,t)}}{\sqrt{\rho(\vec{x}_1,\vec{x}_2,...,t)}}\right)+V(\vec{x}_1,\vec{x}_2,...,t)\right]\ ,
\label{eom}
\end{equation}
where the second term in the brackets generates the classical force and the first encodes a force from the interaction between quantum worlds.  The acceleration can be expressed in terms of the velocity fields via,
\begin{equation}
\vec{a}_k(\vec{x}_1,\vec{x}_2,...,t)=\sum_i \big(\vec{v}_i(\vec{x}_1,\vec{x}_2,...,t)\cdot\vec{\nabla}_i\big)\vec{v}_k(\vec{x}_1,\vec{x}_2,...,t)+\frac{\partial \vec{v}_k(\vec{x}_1,\vec{x}_2,...,t)}{\partial t}\ .
\label{accel}
\end{equation}

To link up with standard discussions of quantum mechanics, one might want to introduce a wave function (a non-fundamental entity) to describe the positions and velocities of the particles---the magnitude giving the density of worlds (when squared) and the phase giving the particle velocities (when the gradient is taken and multiplied by $\frac{\hbar}{m}$, as in \eqref{guideq2}),
\begin{equation}
\Psi(\vec{x}_1,\vec{x}_2,...,t)=\sqrt{\rho(\vec{x}_1,\vec{x}_2,...,t)}e^{i\theta(\vec{x}_1,\vec{x}_2,...,t)}\ 
\label{first}
\end{equation}
\begin{equation}
\vec{v}_k(\vec{x}_1,\vec{x}_2,...,t)=\frac{\hbar}{m_k}\vec{\nabla}_k\theta(\vec{x}_1,\vec{x}_2,...,t)\ ,
\label{guideq3}
\end{equation}
One can only find an angle $\theta$ whose gradients give the velocity fields $\vec{v}_i$ via \eqref{guideq3} if the velocity fields satisfy the \emph{quantization condition} (see \citealp[sec. 7]{takabayasi1952}; \citealp{wallstrom1994}; \citealp[sec. 6]{sebens2015}): Integrating the momenta of the particles along any closed loop in configuration space gives a multiple of Planck's constant, $h=2\pi\hbar$,
\begin{equation}
\oint{\left\{\sum_i{\left[m_i\vec{v}_i \cdot d\vec{\ell}_i\right]}\right\}}=nh\ ,
\label{requirement}
\end{equation}
where the arguments of $\vec{v}_i$ are omitted to simplify notation.  Because the velocity fields are undefined where $\rho=0$ (as there are no worlds to include particles with velocities), this condition only applies to loops which do not pass through---though they may encircle---points where $\rho=0$.  The fact that wave functions which differ by a global phase describe the same state follows immediately from (and is explained by) \eqref{first} and \eqref{guideq3}---the putatively distinct wave functions describe the same world density and velocity fields.

Suppose for simplicity that every world contains just two identical particles. In a given world I label one ``particle 1'' and the other ``particle 2'' and attempt to calculate the motion of particle 1.  To do so I need to use \eqref{eom} which requires as input to the quantum potential the density of worlds $\rho$.  Suppose that in a second world there are two particles at $\vec{y}$ and $\vec{z}$, each near one of those in the first world.  Does the presence of this world contribute positively to $\rho(\vec{y},\vec{z},t)$ or to $\rho(\vec{z},\vec{y},t)$?  This will matter for determining the acceleration of particle 1 via \eqref{eom}.  The answer to the question seems to depend on whether it is particle 1 which is at $\vec{y}$ in the second world and particle 2 at $\vec{z}$ or vice versa.  More precisely, it will depend on whether the particle at $\vec{y}$ in this second world bears a certain connection to the particle labeled 1 or the one labeled 2 in the first world.  Suppose first that there is a fact of the matter about whether in the second world it is particle 1 at $\vec{y}$ and particle 2 at $\vec{z}$ or vice versa; that there is a correspondence between particles in different worlds.  Then we can sensibly imagine densities of worlds $\rho$ and velocity fields $\vec{v}_i$ which are not symmetric and these will be described by wave functions which are neither symmetric nor antisymmetric.  This theory can be called \emph{linked Newtonian quantum mechanics}.  The introduction of such connections between particles in different worlds by linked Newtonian quantum mechanics is unnatural and unnecessary.  \emph{Unlinked Newtonian quantum mechanics} simply omits the connections.  Note that the link under discussion here is distinct from the one considered in the previous section; it is a link between particles across worlds, not a link between particles and the wave function.\footnote{These links are reminiscent of the haecceities which philosophers have introduced to pick out the same object across possible worlds. But, as in this case the quantum worlds are interacting entities not separate possibilities, it's not quite right to think of the links as haecceities.  However, one could say that linked Newtonian quantum mechanics posits ``quasi-haecceities'' or ``quantum haecceities.''}

In unlinked Newtonian quantum mechanics the symmetry dichotomy is a result of the way wave functions are \emph{constructed}: one can allow arbitrary arrangements of the fundamental objects---particles inhabiting different worlds---but the wave function constructed to describe these particles will always be either symmetric or antisymmetric.  The reason asymmetric states are forbidden is that no arrangement of the fundamental objects could give rise to them.  This is in contrast to unlinked Bohmian mechanics where we began with the \emph{constraint} that the various gradients in \eqref{guideqred} must yield the same velocities and showed that the symmetry dichotomy ensures that this will always be so.\footnote{To be fair, unlinked Newtonian quantum mechanics does include its own constraint on allowed states, the quantization condition \eqref{requirement}.  However, unlike the Bohmian constraint, relaxing the quantization condition would not permit asymmetric wave functions, just states in which no wave function at all can properly describe the positions and velocities of the particles.  Further, linked Newtonian quantum mechanics (which allows for asymmetric wave functions) also includes the quantization condition.  For these reasons, I say that it is the way wave functions are constructed---as opposed to a constraint on allowed states---which explains why asymmetric wave functions cannot occur in unlinked Newtonian quantum mechanics.}

\textbf{Step 1: Symmetry of $\bm{R}$}\ \ \   Demonstrating that $R$ must be symmetric is more immediate than in the previous section and presented here as step 1.  According to unlinked Newtonian quantum mechanics there is no correspondence between particles in different worlds; that is, there is no fact of the matter about whether the particle at $\vec{y}$ in the second world (in the example above) corresponds to particle 1 or particle 2 in the first world. If this is the case, we cannot say that the second world contributes to just $\rho(\vec{y},\vec{z},t)$ or $\rho(\vec{z},\vec{y},t)$.  Instead, we should say that it contributes to both.  Each world should be plotted twice in configuration space and $\rho$ introduced to describe the density of these points.  For $N$ particles, each world must be plotted $N!$ times.  A density constructed in this way for multiple particles will automatically be symmetric under the exchange of two particle locations,
\begin{equation}
\rho(\vec{x}_1,\vec{x}_2,...,t)=\rho(\vec{x}_2,\vec{x}_1,...,t)
\ .
\label{rhosym}
\end{equation}
Since $\rho=R^2$ and $R>0$, $R$ must be symmetric.  The density $\rho$ does not merely describe the worlds.  It also determines how they move---in conjunction with the potential $V$ via the force law \eqref{eom}.  This recipe for constructing $\rho$ clarifies how the force law is supposed to function in unlinked Newtonian quantum mechanics (much like unlinked Bohmian mechanics came with a certain understanding of the guidance equation, as discussed in the previous section).  When the density is constructed in this way, the force on a given particle does not to depend on any identification of particles across worlds.

\textbf{Step 2: Symmetry of $\bm{\vec{\nabla}\theta}$}\ \ \  The velocity fields which are generated from the actual particle velocities will satisfy
\begin{align}
\vec{v}_{1}(\vec{x}_1,\vec{x}_2,\vec{x}_3,...,t)&=\vec{v}_2(\vec{x}_2,\vec{x}_1,\vec{x}_3,...,t)
\nonumber
\\
\vec{v}_{2}(\vec{x}_1,\vec{x}_2,\vec{x}_3,...,t)&=\vec{v}_1(\vec{x}_2,\vec{x}_1,\vec{x}_3,...,t)
\nonumber
\\
\vec{v}_{3}(\vec{x}_1,\vec{x}_2,\vec{x}_3,...,t)&=\vec{v}_3(\vec{x}_2,\vec{x}_1,\vec{x}_3,...,t)
\nonumber
\\
&\ \vdots \ \ \ \ \ \ \ \ \  \ \  \  \ \ \ \ \ \ \  \ \ \ \ 
\ .
\label{vsym}
\end{align}
Whether we map a world to $(\vec{x}_1,\vec{x}_2,...)$ or $(\vec{x}_2,\vec{x}_1,...)$ in configuration space, the velocity of the particle at $\vec{x}_1$ will be the same (and similarly for $\vec{x}_2$, etc.).  The condition on the phase of the wave function in \eqref{foralliagreement} follows from \eqref{guideq3} and \eqref{vsym} and holds in exactly the same cases (when $\rho\neq 0$\footnote{For the next step we must assume, as in the case of Bohmian mechanics, that this region is connected (or at least, that it is over time---see footnote \ref{potato}).}).

\textbf{Step 3: $\bm{\alpha=n\pi}$}\ \ \   From \eqref{foralliagreement} it follows, as before, that the values of the wave function at points in configuration space related by a permutation differ by a constant phase \eqref{alphaphase2} and, because two permutations returns you to the original point, that the wave function must be either symmetric or antisymmetric under this permutation.  The same generalization applies: a multi-particle wave function will be either symmetric under all permutations or antisymmetric under all permutations.

The explanation of the symmetry dichotomy in Newtonian quantum mechanics can be summarized as follows:  At the fundamental level there are just particles whizzing about in space (satisfying \eqref{requirement}).  These particles are grouped into a number of worlds, each of which has some quarks, electrons, photons, etc. in it.  When we represent this collection of worlds in terms of a density and a set of velocity fields on configuration space, we introduce a redundancy.  To properly fill up the configuration space, each world must appear multiple times.  The restriction on states which emerges at the non-fundamental level (the symmetry dichotomy) is a result of the redundancy introduced by this choice of representational mechanism.

\begin{figure}[htb]
\center{\includegraphics[width=7 cm]{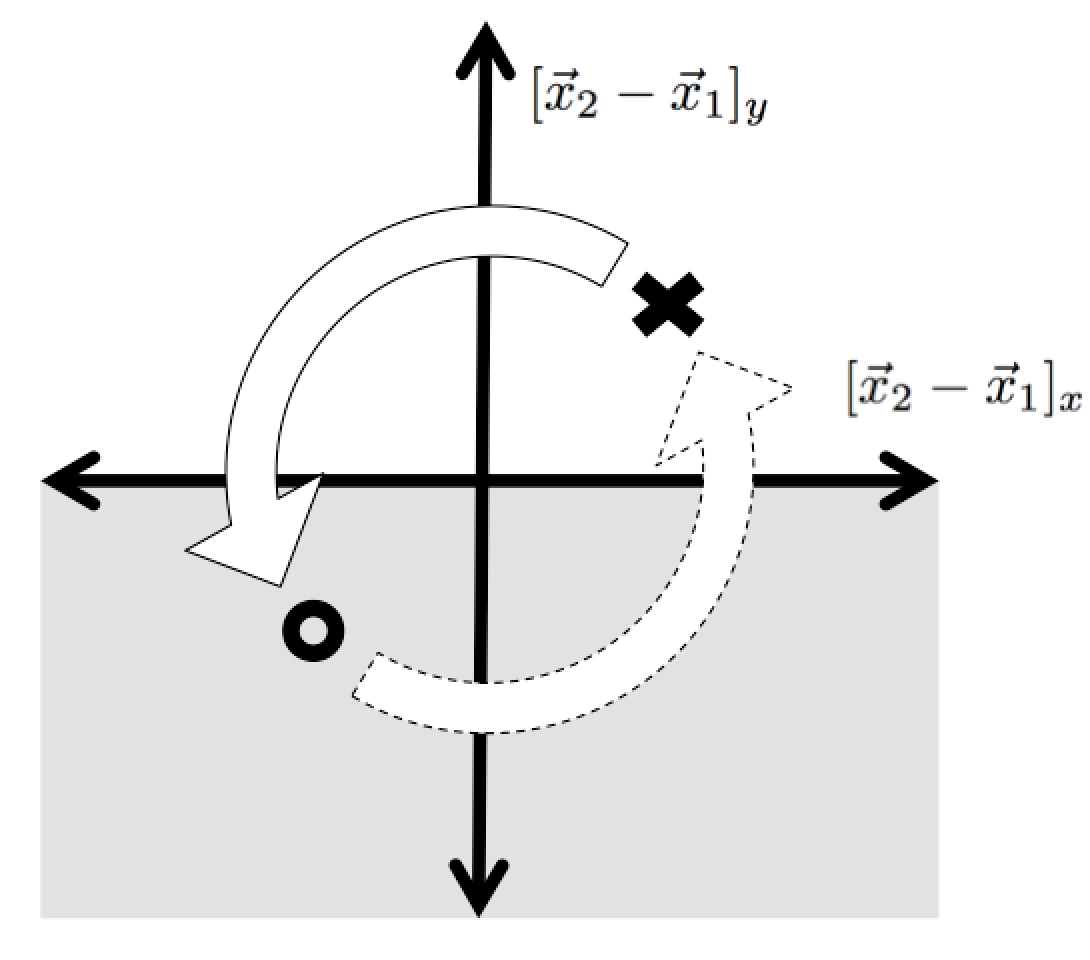}}
\caption{Paths through the space of relative configurations of two particles in two spatial dimensions.}
  \label{2P2D}
\end{figure}

By constraining the behavior of identical particles the quantization condition \eqref{requirement} reveals an important aspect of its strength.  In \citet[\textsection 6]{sebens2015} I gave an example of how the quantization condition leads to the quantization of angular momentum.  It also plays a role in explaining why permuting the locations of two particles yields a phase shift of $n\pi$.\footnote{In the previous argument it played the role of ensuring that the velocity fields can be expressed as gradients of a phase, via \eqref{guideq3}, which was needed to move from \eqref{vsym} to \eqref{foralliagreement}.}  Consider the possible configurations of two identical particles in two dimensional space.  We can describe these particles in terms of the center of mass coordinate $\vec{x}_{cm}=\frac{\vec{x}_1+\vec{x}_2}{2}$ and the relative coordinate $\vec{x}_{rel} = \vec{x}_2-\vec{x}_1$.  In these coordinates, the quantization condition becomes,
\begin{equation}
\oint{m\left(2 \vec{v}_{CM} \cdot d\vec{\ell}_{CM}+\frac{1}{2}\vec{v}_{rel} \cdot d\vec{\ell}_{rel}\right)}=nh\ ,
\label{newquant}
\end{equation}
where $\vec{v}_{CM} = \frac{\vec{v}_1+\vec{v}_2}{2}$ and $\vec{v}_{rel} = \vec{v}_2-\vec{v}_1$.  The relative coordinates are plotted in figure \ref{2P2D} for fixed center of mass; the center point represents coincidence and points equally and oppositely displaced from the origin correspond to the same arrangement of particles.  Consider the path through configuration space shown in figure \ref{2P2D} which starts at the point marked by a cross and ends at the circle.  These marked points are related to each other by permutation of the particle positions.  As the path is not a closed loop in configuration space, it might initially appear that the integral of the relative momentum $m \vec{v}_{rel} = 2\hbar\:\vec{\nabla}_{\vec{x}_{rel}}\theta $ along the path is not constrained.  However, the integral along the dotted path must be the same as that on the first path (since we are taking a second trip through the same worlds; the direction is reversed but so is $\vec{v}_{rel}$ since $\vec{v}_1\leftrightarrow\vec{v}_2$ by \eqref{vsym}) and the closed loop formed by both paths must result in an integer multiple of $2 h$ by \eqref{newquant}.  Thus the integral of the relative momentum along first path must yield a multiple of $h$ and the integral of $\vec{\nabla}_{\vec{x}_{rel}}\theta$ along this path must yield a multiple of $\pi$.\footnote{\citet{takabayasi1952} imposes a version of this as a second quantization condition---requiring fermionic wave functions to yield an odd multiple of $h$, in his eq. 7.6---in addition to the quantization condition discussed above, \eqref{requirement}.}  The quantization condition is odd and the fact that the explanation of the symmetry dichotomy relies on it may be unsatisfying as the condition itself has yet to be explained.  The point here is just that no further symmetrization postulate is needed in addition to the quantization condition to arrive at the symmetry dichotomy.

Figure \ref{2P2D} may help us understand the origin of the Pauli exclusion principle in Newtonian quantum mechanics.  If the particles are fermions, the integral of the relative momentum along the loop must be non-zero.  Further, it must be constant as we consider contracting the loop (provided we don't cross a point where $R=0$).  Since the length of the loop decreases, the particle velocities must increase.  The point of coincidence is in ``the eye of the hurricane'' and the density of worlds must drop to zero since the relative velocity $\vec{v}_{rel}$ cannot be defined at the origin.

An important lesson of the discussion in this section and the one before is that on neither the Bohmian nor the Newtonian picture is the source of the symmetry dichotomy the fact that multiple particles of the same type are in some heavy sense ``identical'' beyond being just ``of the same type.''  For the Bohmian, it is not about the identity of the particles with each other,\footnote{\citet[sec. 7.1.4]{holland} makes this point nicely.} but the connection between the particles and the dimensions of the configuration space on which the wave function is defined.  For the Newtonian, it is not about the identity of particles within a world, but the connection between particles in different worlds.

\section{Explaining the Dichotomy using Reduced Configuration Space}\label{reduced}

There is something odd about the full configuration space in the previous sections: the same arrangement of particles is represented by multiple distinct points in the space.  This redundancy can be removed by moving to a reduced configuration space where each distinct arrangement appears once and only once (discussed in \citet{laidlaw1971,leinaas1977} and in Bohmian mechanics by \citet{sjoqvist1995, brown1999, durr2006, durr2007, bmtextbook, goldstein2014}).  There are a variety of ways to mathematically formulate quantum mechanics on reduced configuration space; here I adopt what I take to be the most straightforward method, using multi-valued wave functions.\footnote{\citet{durr2006, durr2007, bmtextbook, goldstein2014} instead prefer to use single-valued wave functions on a universal covering space of the reduced configuration space.}  Reduced configuration space provides a new arena in which to rerun what are essentially the same arguments as before.  Doing so is worthwhile for seeing that the same assumptions are needed and for setting up the discussion of metaphysics in the next section.

\begin{figure}[p!]
\center{\includegraphics[width=7 cm]{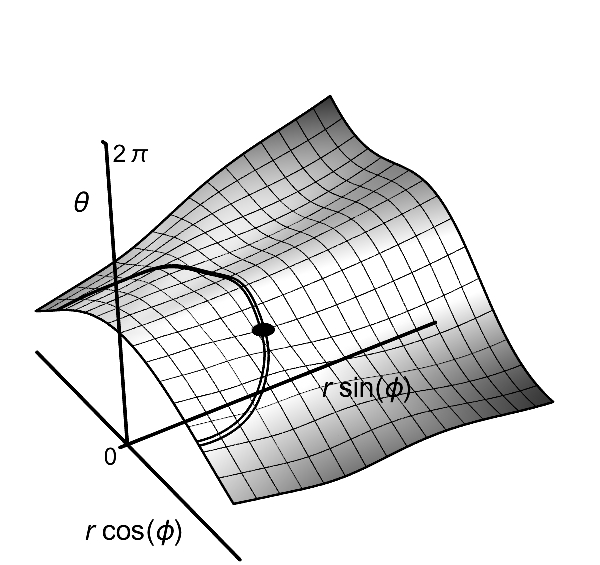}\includegraphics[width=7 cm]{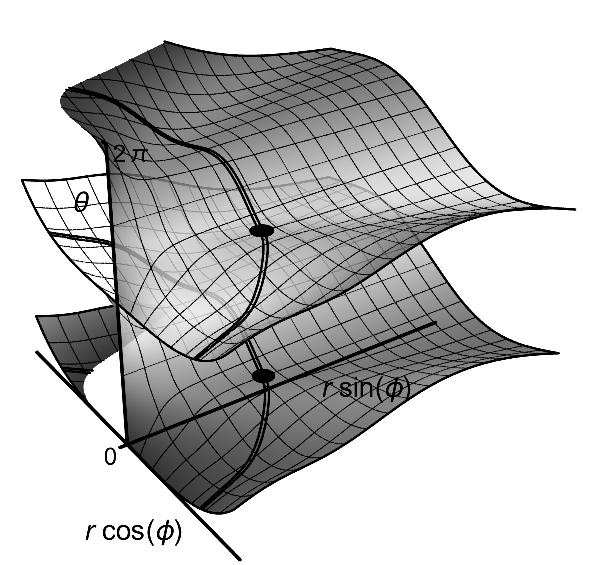}\\(a) symmetric \hspace{4 cm} (b) antisymmetric\\\includegraphics[width=7 cm]{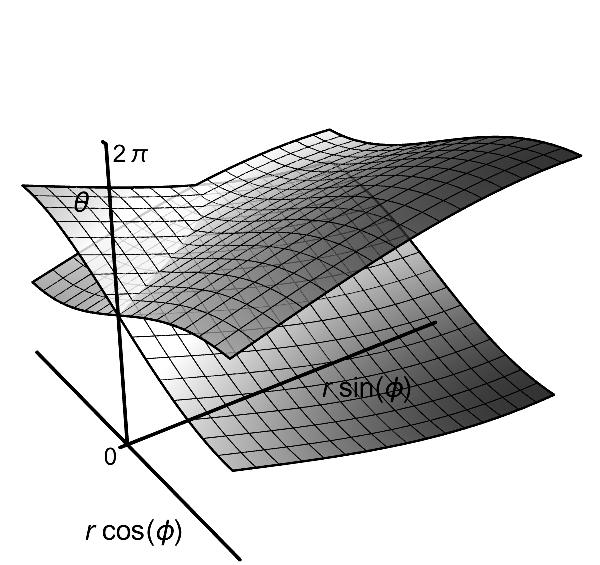}\\(c) asymmetric}
\caption{Here are three plots of possible wave function phases in the reduced relative configuration space for two particles in two dimensional space.  The first plot depicts a single-valued symmetric wave function.  The second depicts a double-valued antisymmetric wave function.  Note that the two branches have the same gradient $\vec{\nabla}_{\vec{x}_{R}}$ everywhere.  In the second figure a corner of the top branch is cut off as it passes $2\pi$ and reappears at the bottom of the plot.  The phase is undefined at $r=0$ where $|\Psi|=0$.  The third figure depicts a wave function which is neither symmetric nor antisymmetric.  There are two branches whose gradients generally disagree.  The two branches happen to cross each other.  Imagine rotating one of the branches $180^{\circ}$ around the $\theta$-axis to picture the wave function over $\vec{x}_{rel}$ in the full configuration space.  In the first two figures I have included the paths discussed in step 2 as double-lines going from an arbitrary point in the reduced configuration space back to itself.}
  \label{dvwf}
\end{figure}

We are free to express the dynamical evolution of a wave function for two identical particles using center of mass and relative coordinates,  $\vec{x}_{CM}=\frac{\vec{x}_1+\vec{x}_2}{2}$ and $\vec{x}_{rel}=\vec{x}_2-\vec{x}_1$.  In these coordinates, the Schr\"{o}dinger equation \eqref{schrod} becomes,
\begin{equation}
i\hbar\frac{\partial}{\partial t}\Psi(\vec{x}_{CM},\vec{x}_{rel},t)
=\left(\frac{-\hbar^2}{2 m}\left(\frac{1}{2}\nabla_{\vec{x}_{CM}}^2+2\nabla_{\vec{x}_{rel}}^2\right)+V(\vec{x}_{CM},\vec{x}_{rel})\right)\Psi(\vec{x}_{CM},\vec{x}_{rel},t)\ .
\label{schrod3}
\end{equation}
Let us now move to the reduced configuration space by identifying points where $\vec{x}_1$ and $\vec{x}_2$ are swapped, that is, by identifying each point $\vec{x}_{rel}$ with $-\vec{x}_{rel}$ and calling it $\vec{x}_{R}$ where the capital $R$ indicates that the coordinate is both ``reduced'' and ``relative.''  In two dimensions this can be achieved by introducing a coordinate $r$ giving the separation between the two particles and $\phi$ specifying the angle the line connecting the two particles makes with the $x$-axis, ranging from $0$ to $\pi$.  The point $\vec{x}_{R}=(r,\phi)$ corresponds to both \{$[\vec{x}_{rel}]_x=r \cos \phi$ and $[\vec{x}_{rel}]_y=r \sin \phi$\} and  \{$[\vec{x}_{rel}]_x= - r \cos \phi$ and $[\vec{x}_{rel}]_y= - r \sin \phi$\}  (points in the gray and white regions of figure \ref{2P2D} are identified).  As two points in the full configuration space are mapped to one in the reduced configuration space, there will in general be two wave function values at a given point $(r,\phi)$ (for $N$ particles, there would be $N!$ values).  We can introduce a single function $\Psi(\vec{x}_{CM},\vec{x}_{R},t)$ to describe the system provided we allow it to be a double-valued function.  The phase of a general double-valued wave function on this space is shown in figure \ref{dvwf}.c.  As is apparent from the figure, the wave function has two parts, branches,\footnote{Though this should be clear from context, let me be explicit: these are branches in the sense of the mathematical analysis of multi-valued functions; they are not the branches of the wave function which are taken to correspond to worlds in the (Everettian) many-worlds interpretation.} on each of which the magnitude and phase of $\Psi$ may potentially independently vary.  Each branch obeys the same dynamic law---the Schr\"{o}dinger equation \eqref{schrod3} with $\vec{x}_{rel}$ replaced by $\vec{x}_{R}$.

As we will see in step 2 of the derivation below, the constraint that the wave function in full configuration space must be single-valued (which I will consider relaxing) restricts the allowed multi-valued wave functions in the reduced configuration space.  However, any wave function on the full configuration space can be represented as a multi-valued wave function on the reduced configuration space.  Thus, it is not the use of the reduced configuration space alone that generates the symmetry dichotomy.  Still, we can derive the symmetry dichotomy in \emph{unlinked} Bohmian mechanics using reduced configuration space \emph{along with} a slight variant of the argument from section \ref{explainingBM}.

\textbf{Step 1: Symmetry of $\bm{\vec{\nabla}\theta}$}\ \ \ The wave function for two identical particles can be written in polar form as $\Psi(\vec{x}_{CM},\vec{x}_{R},t)=R(\vec{x}_{CM},\vec{x}_{R},t)e^{i\theta(\vec{x}_{CM},\vec{x}_{R},t)}$, where $R$ and $\theta$ may be multi-valued.  The particle velocities are given by the guidance equation,
\begin{align}
\vec{v}_{CM}(t)&=\frac{\hbar}{2m}\vec{\nabla}_{\vec{x}_{CM}}\theta(\vec{x}_{CM},\vec{x}_{R},t)
\nonumber
\\
\vec{v}_{R}(t)&=\frac{2\hbar}{m}\vec{\nabla}_{\vec{x}_{R}}\theta(\vec{x}_{CM},\vec{x}_{R},t)\ ,
\label{}
\end{align}
and only well-defined if each branch of the wave function yields the same velocities.  Thus $\vec{\nabla}_{\vec{x}_{CM}}\theta$ and $\vec{\nabla}_{\vec{x}_{R}}\theta$ must have the same value on each branch at every point in the reduced configuration space (as is the case in figure \ref{dvwf}.a and \ref{dvwf}.b but not \ref{dvwf}.c).  As in section \ref{explainingBM}, we must assume that this agreement holds for merely possible particle configurations, not just the actual one (this can be given the same temporally non-local justification as before).

\textbf{Step 2: $\bm{\alpha=n\pi}$}\ \ \    At every point in the reduced configuration space where they are defined, the gradients of the phase are single-valued.  Because the gradients of $\theta$ are single-valued, the path integral of the gradients of $\theta$ along a curve (giving the change of $\theta$ along the curve) is invariant under deformations of the path (keeping the endpoints fixed) that don't cross a point where $R=0$ (where the gradients of $\theta$ are undefined).  If there are no points where $R=0$, $\theta$ is single-valued.  If there are points where $R=0$, $\theta$ may be multi-valued (as in figure \ref{dvwf}.b).  For example, $R$ may be zero at the points of the reduced configuration space where the particles coincide, where $\vec{x}_{R}=0$.  Call these points the ``region of coincidence.''  If the two particles are moving in one dimensional space, the region of coincidence is a hyperplane which separates the reduced configuration space into two disconnected regions. (A hyperplane is a subspace of dimension one less than the original space.  In the case of two particles in one spatial dimension, the region of coincidence is a one dimensional subspace of the two dimensional reduced configuration space.)  As in section \ref{explainingBM}, the symmetry dichotomy cannot be derived if the $R\neq 0$ region is disconnected---which it would be if $R=0$ in the region of coincidence.

If the particles are moving in two dimensions the region of coincidence is a two dimensional subspace of the four dimensional reduced configuration space.  The region of coincidence can helpfully be thought of as a ``hyperline'' since it has two fewer dimensions than the reduced configuration space.  Consider a path from an arbitrary point in the reduced configuration space back to itself which encircles the region of coincidence once (e.g. the path from the cross to the circle in figure \ref{2P2D}, which becomes a closed curve when represented in the reduced configuration space as in figures \ref{dvwf}.a and \ref{dvwf}.b; note that in \ref{dvwf}.b two values are assigned to each point along this path because the phase is double-valued).  This curve cannot be continuously deformed to a point without crossing the region of coincidence and thus the integral of the gradients of $\theta$ along it may be non-zero.  Now consider the same path repeated twice.  The new longer path is the image of a closed path on the full configuration space (e.g. the solid and dotted paths in figure \ref{2P2D} taken together).  Since the wave function must be single-valued on the full configuration space, this integral must yield a phase difference of $2\pi n$.  But, it must also be twice the integral along the original path.  Thus the original integral must be $0$ or $\pi$ (in figure \ref{dvwf}.a following the original path gives a net contribution of $0$ to $\theta$ whereas in \ref{dvwf}.b there is a contribution of $\pi$ from circling the origin once).  So, $\theta$ must either have a single value or two values that differ by $\pi$ at the original point.\footnote{The quick derivation of $\alpha=n\pi$ given in section \ref{explainingBM} cannot be repeated since two points in \eqref{alphaphase2} have been identified.  Instead we need to integrate the phase along curves, as in footnote \ref{earliermess}.}

If the particles are moving in three dimensions the region of coincidence is a three dimensional subspace of a six dimensional space (a ``hyperpoint'' since the difference in dimensions is three).  In this situation the previous argument is equally applicable but another is available as well, one which does not assume that the wave function is single-valued on the full configuration space.  Again consider a path from an arbitrary point in the reduced configuration space back to itself which encircles the region of coincidence once, repeated twice.  It turns out that such a path can be continuously deformed to a point without crossing the singularity (though the single encircling of the singularity cannot).  For details of how the deformation is done, see \citet[pg. 9]{leinaas1977} (they employ this topological insight in a different way, not considering particle interpretations or integrating the gradients of the phase).  It should be noted that this trick fails if there is a sufficiently troublesome set of additional points where $R=0$ limiting one's ability to deform the curve; the appeal to single-valuedness of the wave function on the full configuration space will generally still be legitimate (both methods fail, as discussed in section \ref{explainingBM}, if the configuration space is broken into disconnected pieces by regions where $R=0$).  This trick is by no means only available in the reduced configuration space.  If we consider, for example, a path of constant $\vec{x}_{CM}$ encircling the origin in the full relative configuration space the above deformation will look like taking this loop and trivially deforming it continuously to a point by going around the point of coincidence at the origin (a move which cannot be done if the particles are moving in two dimensional space).  Just as in ordinary space a curve encircling a line cannot be contracted to a point without crossing it but a curve encircling a point can, in the full configuration space a curve encircling the ``hyperline'' of coincidence (in the two dimensional case) cannot be contracted to a point but a curve encircling the ``hyperpoint'' (in the three dimensional case) can.\footnote{A curve encircling the ``hyperpoint'' once in \emph{reduced} configuration space cannot be contracted to a point because of the more complex structure of the space \citep[pg. 9]{leinaas1977}.}  This provides a proof that anyons are forbidden in three or more dimensions (even if multi-valued wave functions are allowed) which does not require the use of reduced configuration space (in contrast to the topological approach of \citealp{leinaas1977}).

\textbf{Step 3: Symmetry of $\bm{R}$}\ \ \  As in section \ref{explainingBM}, the single-valuedness of $R$ can be argued for either by requiring that $R^2$ straightforwardly yield probabilities or by arguing that a double-valued $R$ will cause the gradients of the phase to become double-valued and the velocities ill-defined.

A single-valued wave function on the reduced configuration space corresponds to a symmetric wave function on the full configuration space.  A double-valued wave function on the reduced configuration space for which the amplitude of the wave function is the same on both branches though the phases always differ by $\pi$ corresponds to an antisymmetric wave function on the full configuration space.  As these are the only possibilities allowed by the above argument, the symmetry dichotomy has been established.\footnote{The above argument is somewhat similar to the one given by (\citeauthor{durr2006}, \citeyear[\textsection 5]{durr2006}, \citeyear[\textsection 1]{durr2007}, \citealp[\textsection 8.5]{bmtextbook}).  They require that there be unique well-defined Bohmian velocities throughout reduced configuration space.  I take unlinked Bohmian mechanics together with the story about troublesome time evolution (in section \ref{explainingBM}) to ground this important assumption.  Here I've sought to connect their sort of approach to \citetalias{bacciagaluppi2003} \citeyearpar{bacciagaluppi2003}.  D\"{u}rr \emph{et al.} use advanced mathematics which, though certainly compact, is (in my opinion) more difficult to understand physically than the version of the story presented in this section.}

Here and in section \ref{explainingBM} we arrived at a restriction on allowed wave functions from the fact that the Bohmian velocities generated from the wave function must be well-defined.  Although I think the explanation of the symmetry dichotomy is essentially the same in reduced configuration space, I'll argue in the next section that the option of running the argument in this alternative space is relevant to subtle ontological questions about Bohmian mechanics.

In the previous section we saw that in unlinked Newtonian quantum mechanics each world is equally well represented by multiple points in the full configuration space.  This redundancy was central to explaining the symmetry dichotomy.  Reduced configuration space may thus appear a more natural setting in which to present the theory: each world will only appear once.  The density of worlds $\rho$ and the velocity fields $\vec{v}_{CM}$ and $\vec{v}_{R}$ will describe this collection of worlds at a coarse-grained level (as before).  These worlds can be described by a wave function such that
\begin{align}
\Psi(\vec{x}_{CM},\vec{x}_{R},t)&=\sqrt{\rho(\vec{x}_{CM},\vec{x}_{R},t)}e^{i\theta(\vec{x}_{CM},\vec{x}_{R},t)}
\nonumber
\\
\vec{v}_{CM}(\vec{x}_{CM},\vec{x}_{R},t)&=\frac{\hbar}{2 m}\vec{\nabla}_{\vec{x}_{CM}}\theta(\vec{x}_{CM},\vec{x}_{R},t)
\nonumber
\\
\vec{v}_{R}(\vec{x}_{CM},\vec{x}_{R},t)&=\frac{2\hbar}{m}\vec{\nabla}_{\vec{x}_{R}}\theta(\vec{x}_{CM},\vec{x}_{R},t)\ .
\label{redconNQM}
\end{align}

\textbf{Step 1: Symmetry of $\bm{R}$}\ \ \  There is a single density of worlds $\rho$ on the reduced configuration space and thus $R$ is single-valued ($\rho=R^2$).

\textbf{Step 2: Symmetry of $\bm{\vec{\nabla}\theta}$}\ \ \ Similarly, the velocity fields on the reduced configuration space will be single-valued.  From \eqref{redconNQM} it follows that the gradients of the phase of the wave function describing the worlds, $\vec{\nabla}_{\vec{x}_{CM}}\theta$ and $\vec{\nabla}_{\vec{x}_{R}}\theta$, will be single-valued as well---provided that there is such a function $\theta$ whose gradients yield both $\vec{v}_{CM}$ and $\vec{v}_{R}$.  In the full configuration space the existence of such a function was ensured by the fact that the velocities satisfy the quantization condition \eqref{requirement}.  In the reduced configuration space, the condition becomes
\begin{equation}
\oint{m\left(2 \vec{v}_{CM} \cdot d\vec{\ell}_{CM}+\frac{1}{2}\vec{v}_{R} \cdot d\vec{\ell}_{R}\right)}=nh\ ,
\label{newquant2}
\end{equation}
as in \eqref{newquant}.  However, this condition need only be satisfied for paths that form closed loops both in the reduced configuration space and the full configuration space.  There are closed paths in the reduced configuration space---like the solid arrow in figure \ref{2P2D}---which are not closed in the full configuration space (though taking such a path twice over will form a closed path).  A constraint on these paths, that the integral in \eqref{newquant2} yield $\frac{n h}{2}$, thus follows from imposing \eqref{newquant2} on paths closed in the full configuration space.

\textbf{Step 3: $\bm{\alpha=n\pi}$}\ \ \    Since $\vec{\nabla}_{\vec{x}_{CM}}\theta$ and $\vec{\nabla}_{\vec{x}_{R}}\theta$ are well-defined and single-valued everywhere where $\rho=0$, this step proceeds exactly as in Bohmian mechanics.  Again, the wave function must be single-valued or double-valued where the two values differ by a phase shift of $\pi$

As in the case of Bohmian mechanics, the use of reduced configuration space for Newtonian quantum mechanics serves as an alternative arena to run what is essentially the same argument as in section \ref{explainingNM}.

\section{The Symmetry Dichotomy and Ontology}\label{ont}

For the purposes of deriving the symmetry dichotomy, the key departure of unlinked Bohmian mechanics from the linked version was a revision of the guidance equation, \eqref{guideq2} to \eqref{guideqred}.  The motivation for the guidance equation of unlinked Bohmian mechanics was the fact that a certain connection between particles and wave was not available for the equation to appeal to.  Whether or not such a connection is present is a metaphysical question, related to metaphysical questions about the fundamental space on which the dynamics occurs and the ontological status of the wave function.  Some metaphysically explicit versions of Bohmian mechanics that have been proposed include such a connection and others do not.

One radical proposal is that the wave function is a field on configuration space guiding a single ``marvelous point'' or ``world-particle'' moving through configuration space \citep{albert1996, albert2013}.  On this picture it is the motion of this point in high dimensional configuration space which is fundamental, and the appearance of particles in three dimensional space is to be explained as resulting from the motion of this point.  The connection between particle and wave appealed to in linked Bohmian mechanics is present: the motion of particle 1 is just the motion of the world-particle in the first three dimensions of configuration space, dimensions which we can straightforwardly take the gradient of the phase with respect to in order to calculate a velocity for that particle.  One could instead take the reduced configuration space, as opposed to the full configuration space, as fundamental in which case unlinked Bohmian mechanics is more natural (it would be strange, but not unthinkable, to have the world-particle associated with both a point in the reduced configuration space \emph{and} a particular branch of the wave function).

Alternatively, one might think that, at the fundamental level, there are both particles moving around in familiar three dimensional space and a wave function living in a separate high dimensional space.  These two spaces may or may not be linked in the relevant sense---that is, there may or may not be a correspondence between certain particles and certain dimensions of the high dimensional space on which the wave function lives.

A third set of options takes three dimensional space alone as fundamental.  The wave function might be a multi-field, a law of nature, or a property encoding the dispositions for Bohmian particles to move in certain ways \citep{belot2012}.  The ``multi'' in ``multi-field'' refers to the fact that the field takes multiple particle positions as input, not that it spits out multiple outputs (like the multi-valued functions in section \ref{reduced}).  If the multi-field takes an ordered list of the locations of each particle as input, we have the connection needed for linked Bohmian mechanics.  If the inputs of the multi-field are unordered (when the particles are identical) only unlinked Bohmian mechanics is possible (but the multi-field must be allowed to be multi-valued if antisymmetric wave functions are to be permitted). If the wave function is a law or property, it could determine velocities either from ordered or unordered lists of particle positions---yielding linked or unlinked versions of the theory.

If we take the wave function to be a \emph{Humean} law (as in \citealp{miller2014, callender2015}), there is a sense in which the way the guidance equation works is to be derived from the particle trajectories and not vice versa.  If the trajectories are sufficiently orderly as to obey \eqref{guideqred} (extended to all particles) and not just \eqref{guideq2} (for some wave function), then the wave function which is introduced to describe the particle motions will be one which satisfies the symmetry dichotomy.  This turns out to allow for a simpler set of laws than general Bohmian trajectories would---such a wave function only needs to be specified on $\frac{1}{N!}$ of the configuration space and then its values everywhere else are fixed by its symmetry(antisymmetry).

The lesson I want to draw for ontology is that, as far as explaining the symmetry dichotomy goes, the various unlinked versions of the theory are superior.  \citet{chen} insightfully recognized that one way to evaluate options for the ontology of Bohmian mechanics is by how well they ground explanations of the symmetry dichotomy and the analysis here has arrived at similar conclusions about which variations of Bohmian mechanics best explain the dichotomy.

I should mention that there is also the possibility of mongrel theories with linked metaphysics and unlinked dynamics.  In this case the ontology would include the connections between particles and wave function necessary for linked Bohmian mechanics but the dynamics would not make use of them.  As it was the dynamics of unlinked Bohmian mechanics---in particular, the guidance equation---which was used to derive the symmetry dichotomy, one might hope to defend these linked ontologies by positing unlinked dynamics.  Though such mongrel theories are consistent, they have the disadvantage of positing unnecessary metaphysical structure which is irrelevant to the dynamics.

The link which is present in linked Newtonian quantum mechanics is a link between particles in different worlds, specifying which particle in some other world is ``the same one as'' or really ``the one corresponding to'' a particular particle in this world.  As was discussed in section \ref{explainingNM}, if we take the fundamental ontology of the theory to be a collection of particles belonging to different worlds moving around in a single three dimensional space (the second ontological option discussed in \citealp[sec. 11]{sebens2015}) then such a link would be unnatural because it would require positing additional structure beyond the position, velocity, and world of each particle.  There is also a radical ontology for Newtonian quantum mechanics analogous to Albert's proposal for Bohmian mechanics where there are many world-particles moving through a 3$N$ dimensional space (the first option in \citealp[sec. 11]{sebens2015}; see also \citealp[sec. 4]{bostrom2014}).  This ontology would have the structure necessary for a linked version of the theory because each set of three dimensions would give the location of ``the same particle'' in each world.  However, we could instead take the space on which these world-particles move to be the reduced configuration space of section \ref{reduced}, yielding an improved unlinked version of the theory.

To the extent that we cannot decide between between competing interpretations of quantum mechanics through empirical testing,\footnote{Empirical data decisively favors relativistic quantum field theory over non-relativistic quantum mechanics.  Thus, one should also consider prospects for a relativistic extension in evaluating the promise of competing interpretations.} we must rely on comparison via theoretical virtues like simplicity, elegance, and explanatory power to determine which proposals are most promising.  One virtue we can use to rank various interpretations is their ability to explain the symmetry dichotomy.  In this article I've argued for the following ranking:  The best explanation is provided by unlinked Newtonian quantum mechanics (in one of its subtle variations described in the previous paragraph).  The symmetry dichotomy is automatically satisfied because symmetric and antisymmetric wave functions describe all possible states of the fundamental objects.  Unlinked Bohmian mechanics earns a close second place.    The wave function must be either symmetric or antisymmetric if the particle velocities derived from the wave function are to be independent of any identification between dimensions of configuration space and particular particles.  This explanation is ranked lower for two main reasons (discussed in section \ref{explainingBM}).  First, asymmetric wave functions are forbidden in an indirect, temporally non-local way.    Second, the proof that the magnitude of the wave function is symmetric is much more complicated than in Newtonian quantum mechanics.  At the bottom of the ranking are interpretations in which the symmetry dichotomy must be postulated, but for which such a postulate receives no explanation beyond the fact that it is needed to account for the data.  This includes linked versions of Newtonian quantum mechanics and Bohmian mechanics along with interpretations of quantum mechanics that do not include particles following definite trajectories, like the (Everettian) many-worlds interpretation and spontaneous collapse theories.\\\\

\noindent\textbf{Acknowledgements}

\noindent I would like to sincerely thank David Baker, Eddy Keming Chen, Laura Ruetsche, and anonymous reviewers for helpful feedback on drafts of this paper, Adam Caulton and Sheldon Goldstein for useful discussion, and the audience at the So Cal Philosophy of Physics Reading Group---in particular Jeffrey Barrett and Benjamin Feintzeig.

\bibliography{symmetrizationbibfile}
\end{document}